\documentclass[12pt]{iopart}
\usepackage{iopams}  

\usepackage{graphicx}
\usepackage{color}
\usepackage{textcomp}

\newcommand{\extder}{\mathrm{d}} 
\newcommand{\half}{\case{1}{2}}
\newcommand{\quarter}{\case{1}{4}}
\newcommand{\abs}[1]{| #1 |}
\newcommand{\norm}[1]{|| #1 ||}
\newcommand{\curl}{\nabla \times}

\begin{document}

\title[]{Giant vortex states in type I superconductors simulated by Ginzburg-Landau equations}

\author{H. Palonen$^{1,2}$, J. J\"aykk\"a$^3$ and P. Paturi$^1$}

\ead{heikki.palonen@utu.fi}

\address{$^1$Wihuri Physical Laboratory, Department of Physics and 
Astronomy, FIN-20014 University of Turku, Finland} 
\address{$^2$National Doctoral Programme in Nanoscience, Turku, Finland}
\address{$^3$Nordita, KTH Royal Institute of Technology and Stockholm University, Roslagstullbacken 23, 106 91 Stockholm, Sweden}

\date{\today}

\begin{abstract}


The quantization of magnetic flux in superconductors is usually seen as vortices penetrating the sample. While vortices are unstable in bulk type I superconductors, restricting the superconductor causes a variety of vortex structures to appear. We present a systematic study of giant vortex states in type I superconductors obtained by numerically solving the Ginzburg-Landau equations. The size of the vortices is seen to increase with decreasing film thickness. In type I superconductors, giant vortices appear at intermediate thicknesses but they do not form a well-defined vortex lattice. In the thinnest type I films, singly quantized vortices seem to be stabilized by the geometry of the sample instead of an increase in the effective Ginzburg-Landau parameter.

\end{abstract}


\submitto{\JPCM}

\maketitle

\section{Introduction}

A well-known property of the superconductors is the quantization of the magnetic flux which is an essential property for many of the applications of superconductivity. In the typical case of a type II superconductor in the mixed state, the magnetic field penetrates the sample as vortices that form an Abrikosov lattice each vortex carrying one quantum of flux. With the recent advances in imaging techniques vortices can be resolved in greater detail compared to the traditional Bitter decoration. Perhaps the most impressive is the work of Hasegawa \emph{et al.}~\cite{Hasegawa5} in which the internal magnetic field structure of vortices was measured. They report vortices in thin Pb films with single and multiple flux quanta. These exotic vortex structures with more than one flux quantum have also more recently been reported in confined superconducting systems and in thin Pb films that had triangular or quasiperiodic artificial pinning sites~\cite{Cren2,Engbarth1,Silhanek3,Kramer4}. Multiquantum vortices are seen in the experiments despite the fact that a vortex with $n>1$ flux quanta are generally thought to be energetically unfavourable. The magnetic energy of a giant vortex scales as $n^2$ while it scales as $n$ if the flux quanta are separated into singly quantized vortices~\cite{Poole2}. In this paper, we use the convention that a giant vortex is a cylindrically symmetric vortex with more than one flux quantum while the non-symmetric case is called a multivortex.

Tinkham was the first to propose that vortices could be the reason behind the increase in the critical field of thin type I superconductors~\cite{Tinkham2}. Later on Maki showed that the Abrikosov lattice is a stable solution to the Ginzburg-Landau equations for arbitrary $\kappa$ provided that the film thickness is sufficiently small~\cite{Maki5}. The work of Maki was continued by Lasher who predicted that, between thick type I films that are in the intermediate state and thin films that are in the mixed state, there is a transitional area where giant vortices form honeycomb lattices and possibly more complex structures~\cite{Lasher1}. The theoretical work on these giant vortex structures is mostly based on linearized Ginzburg-Landau equations with the exception of a few recent numerical papers~\cite{Schweigert1,Berdiyorov1,Shi9,Sweeney1}. The current methods in numerical modelling allow us to solve the complete Ginzburg-Landau equations to address the problem of the complex giant vortex structures in type-I superconductors.

In this paper we present a systematic study of vortex states in low-$\kappa$ superconducting thin films in low magnetic fields. We have numerically determined the local energy minima of the Ginzburg-Landau functional for a series of thin films of different thicknesses. The results show stable giant vortex configurations at certain thicknesses and a change in the size of vortices with film thickness.

\section{Computational model}

The static Ginzburg-Landau equations were solved by finding a (local) minimum of the associated energy functional. It is convenient to write the energy in a dimensionless form leaving only the overall energy scale dimensional which does not affect the solutions. The dimensionless energy is 
\begin{equation}
  \label{eq:scaled_E}
    E = 
    \int \extder^3 {x}
    \Big(
        \half\norm{(\nabla + i {\vec{A}}){\psi}}^2 +
        \half\norm{\curl{{\vec{A}}}}^2 
        + \quarter \kappa^2 (\abs{\psi}^2-1)^2
    \Big),  \nonumber
\end{equation}
where $\kappa = \sqrt{\beta/(2\mu_0\hbar^2\gamma^2q^2)}$ is the dimensionless Ginzburg-Landau parameter and $\gamma = 1/(4m_\mathrm{e})$ and $q = 2e$. Here $\psi$ is the order parameter, $\vec{A}$ is the vector potential, $\mu_0$ is the permeability of free space, $m_\mathrm{e}$ is the mass of an electron and $\beta$ is the coefficient of the $\abs{\psi}^4$ term as usual in the context of Ginzburg-Landau theory. The penetration depth is absorbed in the overall energy scale and therefore $\lambda = 1$. The method used to minimize \eref{eq:scaled_E}, the used boundary conditions and the details of restricting $\psi$ to model pinning sites are described in our earlier work~\cite{Palonen2}. In short, the boundaries are so that $x$ direction is periodic and the $y$ and $z$ directions end in vacuum with magnetic field $B$ parallel to the $z$. 

The Ginzburg-Landau functional was minimized for $\kappa = 0.5$ and $\kappa = 1.0$ in thin film geometry (27 $\mu$m x 37 $\mu$m x $d$) with thicknesses $d= 3.5\xi, 4\xi, 5\xi, 7\xi, 10\xi$ and $15\xi$. The magnetic field values used were 0.2, 0.3, 0.4, 0.5 and 0.6 mT. Any lengths or magnetic field values that are given in dimensional form in this paper assume that $\lambda =$ 150 nm. The simulations were chained so that for each thickness the result with the lower magnetic field was used as the starting point for the higher field value simulation. The Meissner state was used as the initial configuration in the 0.2 mT simulations. In all the simulations there were two pinning sites at the same positions to break the symmetry caused by periodic $x$ boundaries. Additionally, the sample was enclosed in vacuum extending 4.5$\lambda$ in the non-periodic directions which allows to include the effects of magnetic stray field energy in the simulations. 

\section{Results and discussion}

\begin{figure}
  \begin{center}
    \includegraphics[width=13.6cm]{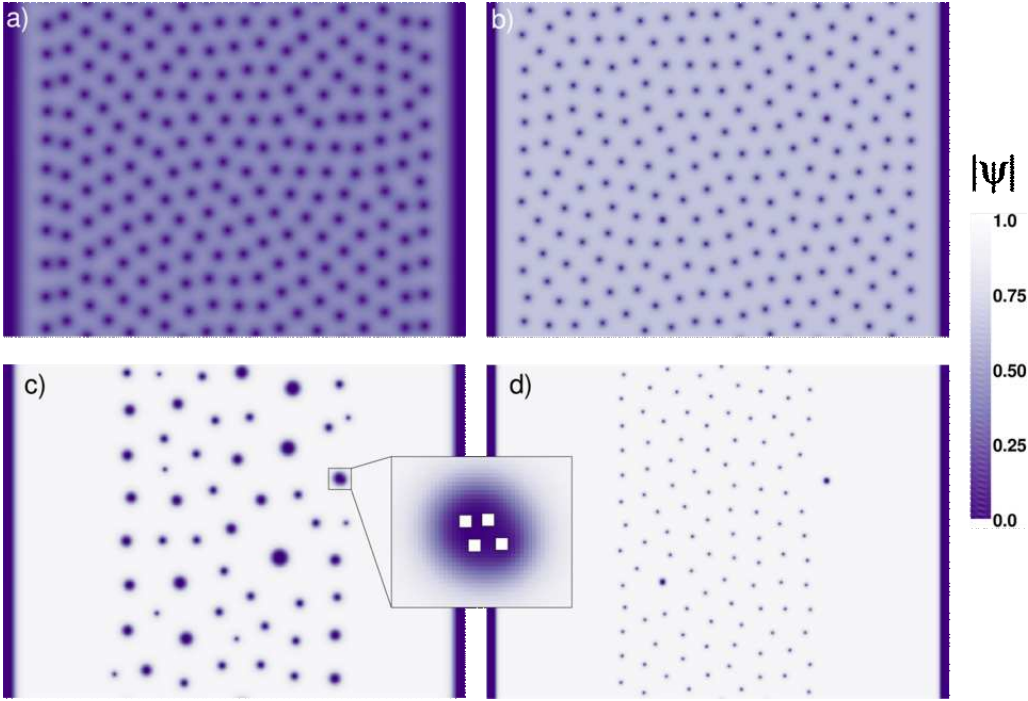}
    \caption{The absolute value of the order parameter shown on a plane sliced through the centre of the sample perpendicular to the magnetic field. The used simulation parameters were a) $\kappa = 0.5$ and $d=3.5\xi$,  b) $\kappa = 1.0$ and $d=3.5\xi$, c) $\kappa = 0.5$ and $d=15\xi$, d) $\kappa = 1.0$ and $d=15\xi$. Magnetic field is 0.5 mT in all the images. The size of the images is 37  $\mu$m $\times$ 27 $\mu$m (830 $\times$ 600 pixels).}
    \label{example}
  \end{center}
\end{figure}

The vortex configurations obtained by the simulations can be seen from the absolute value of the order parameter that is shown in \fref{example} for both $\kappa = 0.5$ and $\kappa = 1.0$ in two cases, the thinnest and the thickest film at $B = 0.5$ mT. The singly quantized vortices in the thinnest films (\fref{example}a and b) are visibly larger in size than in the thick films (\fref{example}c and d) for both values of $\kappa$. In the thick film with $\kappa = 0.5$ the larger vortices are giant vortices with vorticities of all values up to six. All the vortex lattices have a strongly disordered triangular lattice which is partly due to the low value of $\kappa$. As we have shown in our earlier work (ref.~\cite{Palonen2}), the disorder is also increased by the pinning sites that can be seen as the two larger vortices in \fref{example}d. The two pinning sites are present at the same position in all the simulations. The giant vortex in the pinning site of which an enlarged version is also shown in \fref{example}c is actually a multivortex since the cylindrical symmetry of the vortex is broken. A closer look at the phase information reveals the individual vortex cores that are marked with white squares. The distinction between a cylindrically symmetric giant vortex with the vortex cores all exactly centered and a multivortex with the vortex cores very closely packed together is quite subtle. However, all multiply quantized vortices in pinning sites seem to be multivortices while the free vortices are giant vortices. This could possibly explain the two types of vortices seen in the experiments of Hasegawa \emph{et al.}~\cite{Hasegawa5}.

\begin{figure}
\begin{center}
\includegraphics[width=13.6cm]{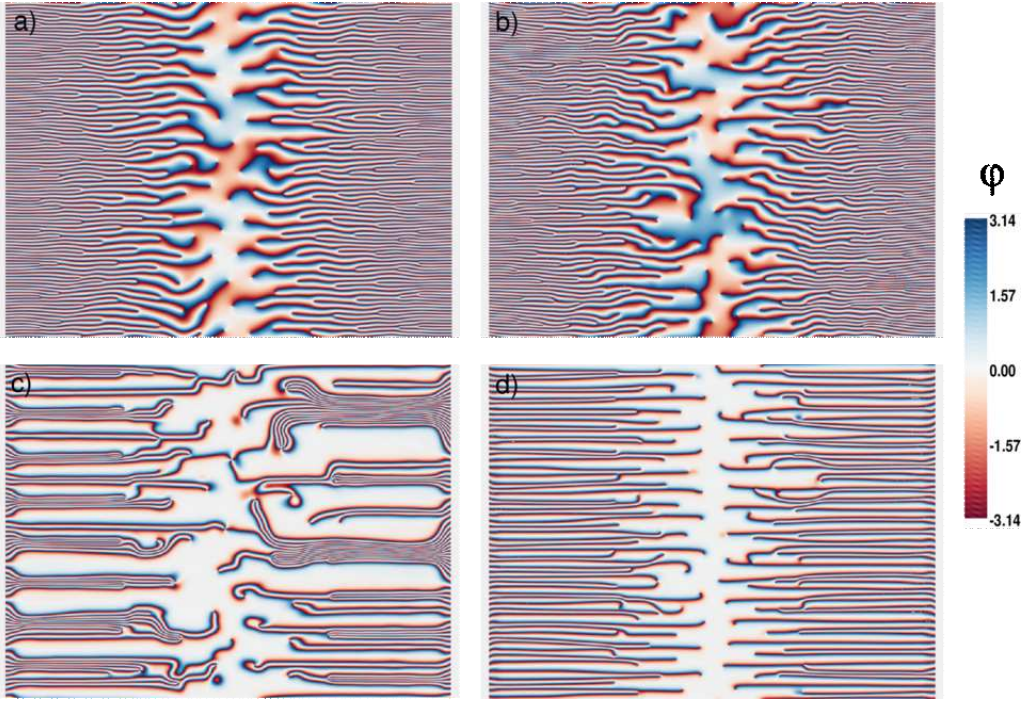}
\end{center}
\caption{The phase of the order parameter shown on a plane sliced through the centre of the sample perpendicular to the magnetic field. The used simulation parameters were a) $\kappa = 0.5$ and $d=3.5\xi$,  b) $\kappa = 1.0$ and $d=3.5\xi$, c) $\kappa = 0.5$ and $d=15\xi$, d) $\kappa = 1.0$ and $d=15\xi$ which are the same parameters as for the images shown in \fref{example}. Magnetic field is 0.5 mT in all the images.}
\label{example_phase}
\end{figure}

The phases of the order parameters corresponding to the absolute values in \fref{example} are shown in \fref{example_phase}. It is clearly visible that the phase contours leading to a giant vortex are clustered together all the way to the sample border. In contrast, the phase contours of the singly quantized vortices are evenly spaced to fill the whole length of the sample. The clustering of the phase contours reflects the nature of the vortex-vortex interaction which is attractive in the case of $\kappa < 1/\sqrt{2}$. The clustering is prominent in the thick film (\fref{example_phase}c) with $\kappa = 0.5$ while the thin film (\fref{example_phase}a) with the same $\kappa$ value shows a similar spread in phase contours as the $\kappa = 1.0$ ones with repulsive vortex-vortex interaction.

\begin{figure}
\begin{center}
\includegraphics[width=6.9cm]{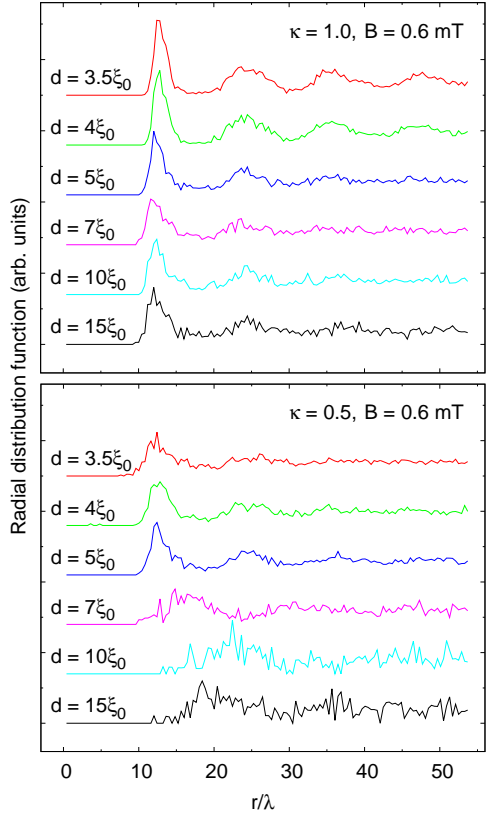}
\caption{The radial distribution functions of the vortex lattices obtained from the simulations for different film thicknesses $d$ with $\kappa = 0.5$ and $\kappa = 1.0$ at 0.6 mT.}
\label{rdf}
\end{center}
\end{figure}

To visualize the order of the vortex lattices we define the two-dimensional case of the radial distribution function as
\begin{equation}
g(r) = \frac{1}{\rho_\mathrm{av}} \frac{\extder n(r+\extder r)}{\extder A(r+\extder r)}
\end{equation}
where $\rho_\mathrm{av}$ is the average density of vortices in the sample, $\extder A$ is the area of the ring from $r$ to $\extder r$ and $\extder n$ is the number of vortices inside it~\cite{Duan2}. The radial distribution functions were calculated from the positions of the vortices in cross-sections similar to the images in \fref{example} and are shown in \fref{rdf}. The vortices in $\kappa = 1.0$ simulations form a better defined vortex lattice than the vortices in the $\kappa = 0.5$ case. In the case of $\kappa = 0.5$, the nearest neighbour peak broadens and moves to larger distances with increasing film thickness. The shift in the peak position is caused by the giant vortices that have larger separation than singly quantized vortices. The broadening of the peak means that the giant vortices do not form a lattice with a well defined nearest neighbour distance. In the case of $\kappa = 1.0$, the thinnest film has stronger peaks than the thickest which indicates more robust vortex lattice in the thin sample.

\begin{figure}
\begin{center}
\includegraphics[width=8.6cm]{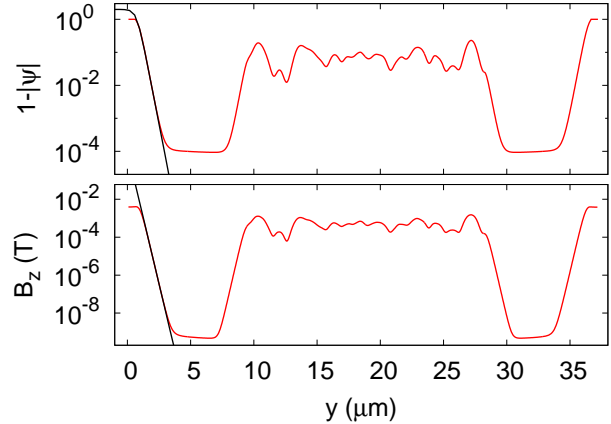}
\caption{The fits (black lines) of $1-\tanh[y/(\sqrt{2}\xi_\mathrm{eff})]$ and $\exp(-y/\lambda_\mathrm{eff})$ to the average profile of the order parameter (top) and magnetic field (bottom). The shown profiles are averages from the case shown in \fref{example}c which was simulated with parameters $\kappa = 0.5$, $d=15\xi$ and $B=0.5$ mT.}
\label{example_fit}
\end{center}
\end{figure}

The change of the vortex size with film thickness was quantified by considering an effective coherence length $\xi_\mathrm{eff}$ and an effective penetration depth $\lambda_\mathrm{eff}$. The effective coherence length was defined by looking at how the order parameter changes near the sample surface, \emph{i.e.} the absolute value of the order parameter should approach zero as $\tanh[y/(\sqrt{2}\xi_\mathrm{eff})]$ and the magnetic field should decrease into the superconductor from the surface as $\exp(-y/\lambda_\mathrm{eff})$~\cite{Poole2}. These were fitted to the average order parameter and field profiles of the samples where the averaging was done to the middlemost splice perpendicular to the magnetic field averaged over the periodic direction. Examples of the fits to the order parameter profile and magnetic field profile are shown in \fref{example_fit}. The fits were limited to the region where the profile is linear in semi-log plot near the sample boundary. The accuracy of the fits is better in the thicker samples where the vortices are not so near the boundary as in the thinner samples. For consistency, these fits were also checked against fitting the single vortex profiles which yielded similar results but proved to be difficult to do systematically.

\begin{figure}
\begin{center}
\includegraphics[width=8.6cm]{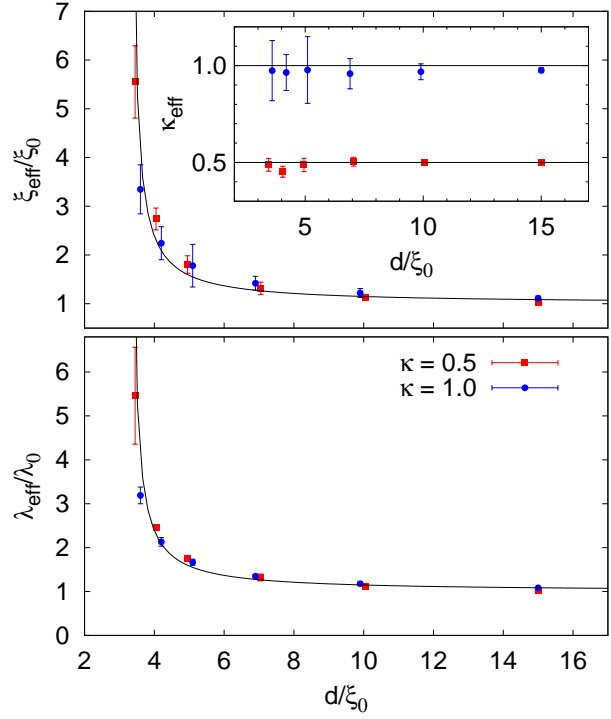}
\caption{The effective characteristic lenghts $\xi_\mathrm{eff}$ and $\lambda_\mathrm{eff}$ derived from the fits in \fref{example_fit} as a function of the film thickness. The black lines show the $1/d$ behaviour as described by the equation~\ref{paksuus}. The inset shows the effective Ginzburg-Landau parameter $\kappa_\mathrm{eff}$ as a function of the film thickness.}
\label{lengths}
\end{center}
\end{figure}

No systematic magnetic field dependence of the fitted characteristic lengths was found. Thus, the fit results for a fixed sample thickness with different magnetic field values were averaged and are shown in \fref{lengths}. The standard deviations of the fit results for the same thickness are indicated with the error bars. The effective lengths $\xi_\mathrm{eff}$ and $\lambda_\mathrm{eff}$, given by the fits, decrease with increasing film thickness. The effective lengths for both $\kappa$-values fall on the same curve when scaled with the bulk values $\lambda_0$ and $\xi_0$. The inset shows how the effective Ginzburg-Landau parameter $\kappa_\mathrm{eff} = \lambda_\mathrm{eff}/\xi_\mathrm{eff}$ is independent of the sample thickness. Therefore, the effective lengths have the same thickness dependence which is shown as the black lines that have the form 
\begin{equation}
\frac{X_\mathrm{eff}}{X_0} = \frac{\xi_0}{d-d_0} + 1, \quad X = \lambda, \xi
\label{paksuus}
\end{equation}
where $d_0 \approx 3.3\xi_0$ is the only free parameter obtained by fitting. For $X = \lambda$ equation~\ref{paksuus} can also be written as 
\begin{equation}
\lambda_\mathrm{eff} =  \frac{\lambda^2_0}{\kappa(d-d_0)} + \lambda_0
\label{pearl}
\end{equation}
where $\kappa = \lambda_0/\xi_0$. Equation~\ref{pearl} is quite similar to the $\Lambda \sim \lambda^2/d$ given by Pearl~\cite{Pearl1} for the interaction length of two Pearl vortices in a thin superconducting film. Both effective lengths diverge at $d_0$ and at thicknesses less than $d_0$ the simulation gives normal state as the minimum energy. The divergence of the effective lengths is obviously unphysical since, for example, Pb thin films are superconducting down to 5 monolayer thicknesses~\cite{Ozer1}. However, the general trend that both effective lengths increase as the thickness decreases is in agreement with experiments that show how the critical field decreases ($H_\mathrm{c} \sim 1/(\xi\lambda)$) as the thickness decreases~\cite{Cody1}. The decrease of the critical field with decreasing thickness continues up to the point when the thickness of the sample starts to restrict the electron mean free path after which the critical field starts to increase~\cite{Cody1,Lischke1}. Since the electron mean free path related effects are not included in our model it is natural that the simulation results start to deviate from the experiments when the mean free path becomes important for the real sample.

It is the stray field outside the sample that is responsible for the change in effective lengths and vortex-vortex interaction with film thickness. This is in good agreement with the work of Brandt in ref.~\cite{5} that shows when taking the stray fields into account the vortex lattice behaves like it had an effective $\kappa$ that tends to infinity as thickness goes to zero. The effective $\kappa$ in Brandt's work is defined through the behaviour of the bulk shear modulus of the vortex lattice which reflects the change in the vortex-vortex interaction to repulsive. In this work $\kappa_\mathrm{eff}$ stays constant as it is defined by the ratio of the effective lengths which differs from Brandt's definition. The result that $\kappa_\mathrm{eff}$ stays constant is interesting, because the vortices seen in thin type I films are explained with electron mean free path increasing $\kappa$ which drives the material into type II behaviour. Our simulations predict that singly quantized vortices could be stabilized just by the geometry of the sample. An alternative viewpoint to effective $\kappa$ is to consider how the expansion of vortices near the surface strengthens the repulsive interaction of vortices which leads to enhanced repulsion in thinner films~\cite{Carneiro1}.

\begin{figure}
\begin{center}
\includegraphics[width=6.9cm]{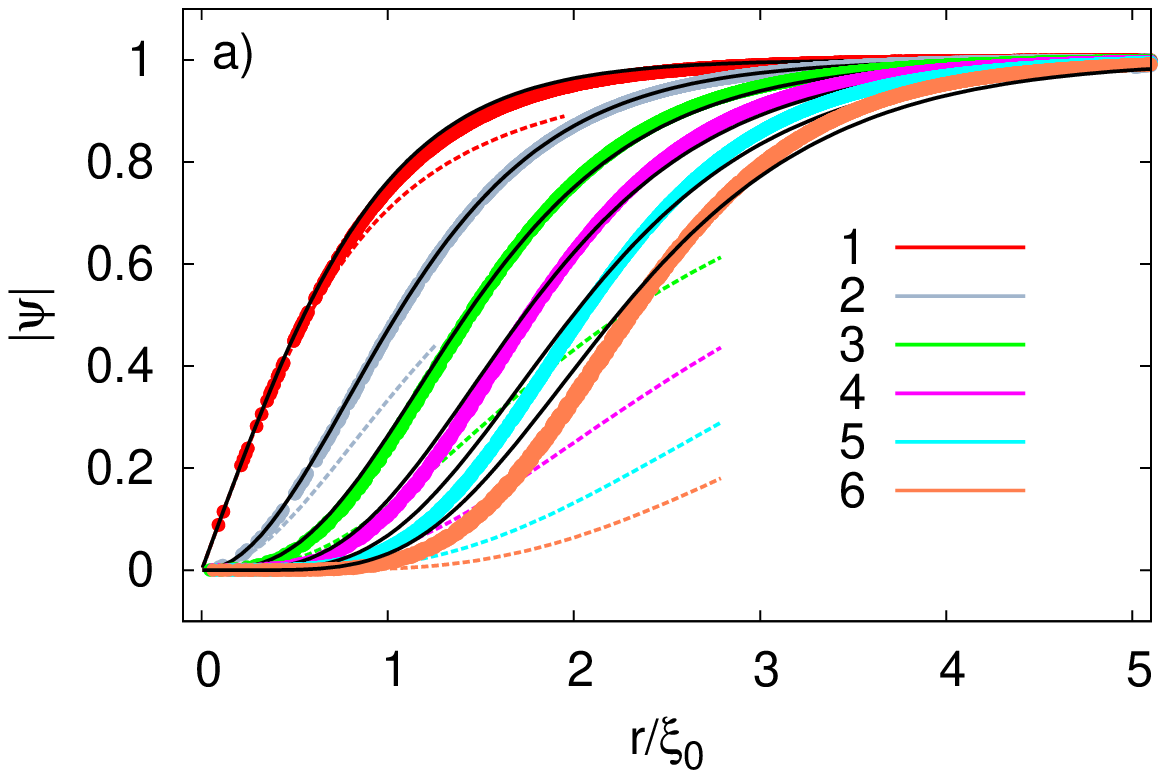}
\includegraphics[width=6.9cm]{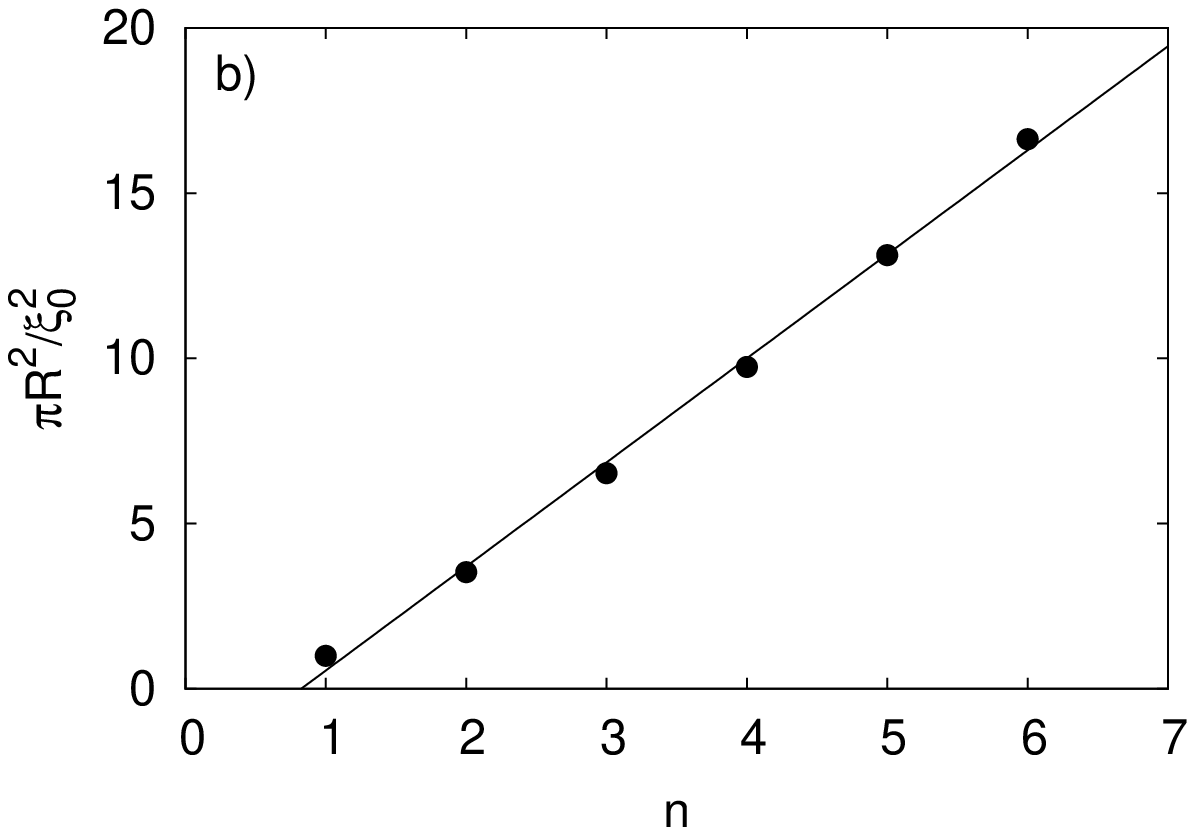}
\caption{(a) The radial profile of the order parameter for flux quanta ranging from one (top) six (bottom). The points show the vortex profiles from our simulation for thickness 15$\xi$ at 5 mT. The Clem model combined with Lasher's (equation~\ref{lasher}) is shown as the dashed lines. The solid black lines show the equation~\ref{omafit}. (b) The area of the cross-section of the vortices as a function of the number of flux quanta. The cross-section is calculated from the area of a circle of radius $R$ that is defined as $\abs{\psi(r=R)}=0.5$. The line is a least squares fit to the points.}
\label{radialprofile}
\end{center}
\end{figure}

Naturally, the giant vortices grow monotonically in size with the increasing number of flux quanta. Examples of the radial profiles of the vortices with different quantum numbers are shown in \fref{radialprofile}a. A practical estimate for the radial shape of a vortex, known as the Clem model~\cite{Clem5}, is given by
\begin{equation}
\abs{\psi_1 (r)} = \frac{r}{\sqrt{r^2+\xi^2_v}},
\label{clem}
\end{equation} 
where $r$ is the radial coordinate and $\xi_v \approx \xi_0 $ is the radius of the vortex. The Clem model gives a rather good estimate of the single vortex shape near the core. The Clem model can be extended to give estimates of multiple quantum vortices with the help of Lasher's equation~\cite{Lasher2} which gives the order parameter of the giant vortex as
\begin{equation}
\psi_n (r) = \left[ \psi_1 \left( \frac{r}{\sqrt{n}} \right) \right]^n
\label{lasher}
\end{equation}
where $n$ is the number of flux quanta and $\psi_1$ is the order parameter of the single vortex. Combining \eref{clem} and \eref{lasher} gives
\begin{equation}
\psi_n = \left[ \frac{r}{\sqrt{r^2 + n\xi^2_0}} \right]^n
\label{combined}
\end{equation}
which is also plotted as the coloured dashed lines in \fref{radialprofile}a for $n= 1\ldots6$. While \eref{combined} catches the general flattening of the vortex core with increasing number of quanta, it deviates quite a lot from the simulation results outside the vortex core. The deviation is not surprising given that the analytical expression is based on linearized Ginzburg-Landau equations and our simulations are quite far from the situation when the linearization is valid. A more practical function that could be used in fitting  experimental data would be
\begin{equation}
\psi_n =  \left[ \tanh\left(\frac{r}{n^{\frac{1}{4}}\xi_0}\right)\right]^n
\label{omafit}
\end{equation}
which is similar to \eref{combined} except that $\psi_1 = \tanh(r/\xi_0)$ and the scaling is changed from $n^{\frac{1}{2}}$ to $n^{\frac{1}{4}}$. These are shown as black lines in \fref{radialprofile}a. The size of the giant vortex increases almost linearly with increasing number of quanta which is shown in \fref{radialprofile}b. The size of the vortex was characterized by the cross-sectional area of the vortex where $\psi < 0.5$. The linear behaviour is to be expected because the magnetic flux density is approximately constant in the vortex core which means that each flux quantum takes up the same amount of space.

\begin{figure}
\begin{center}
\includegraphics[width=6.9cm]{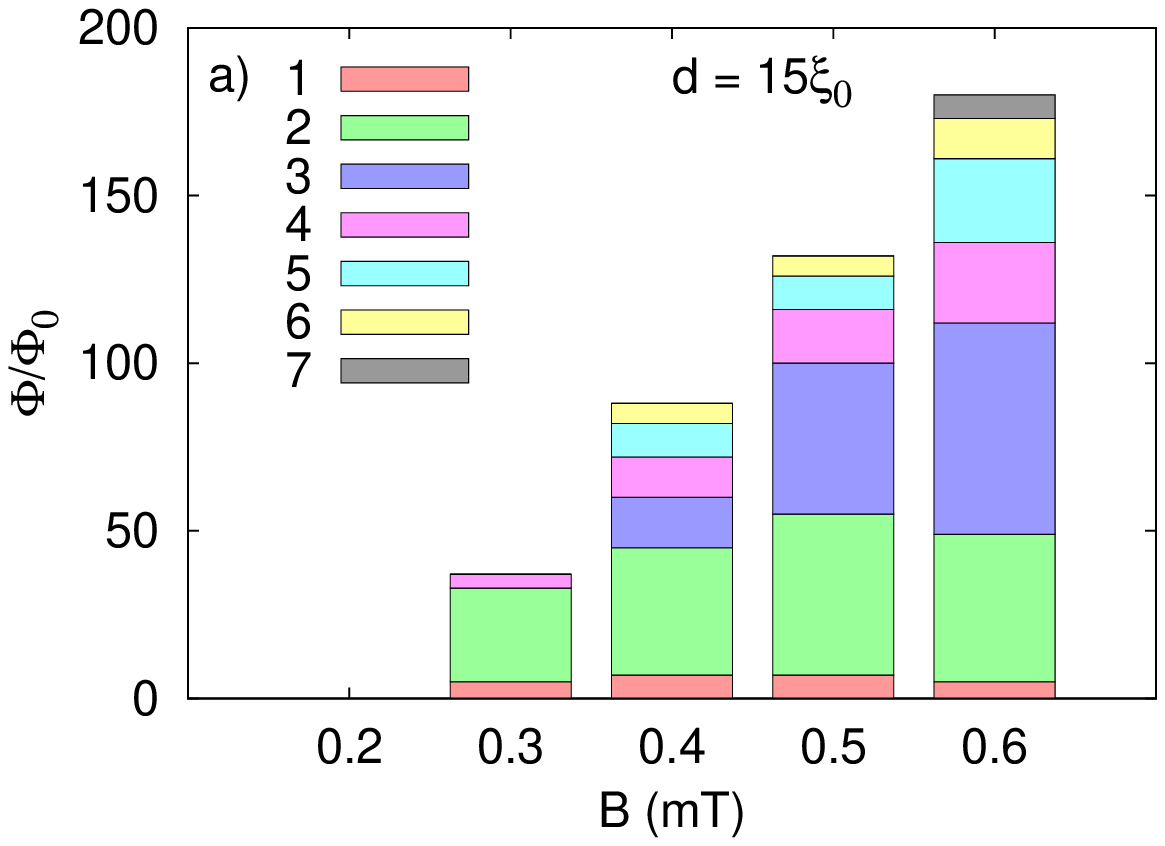}
\includegraphics[width=6.9cm]{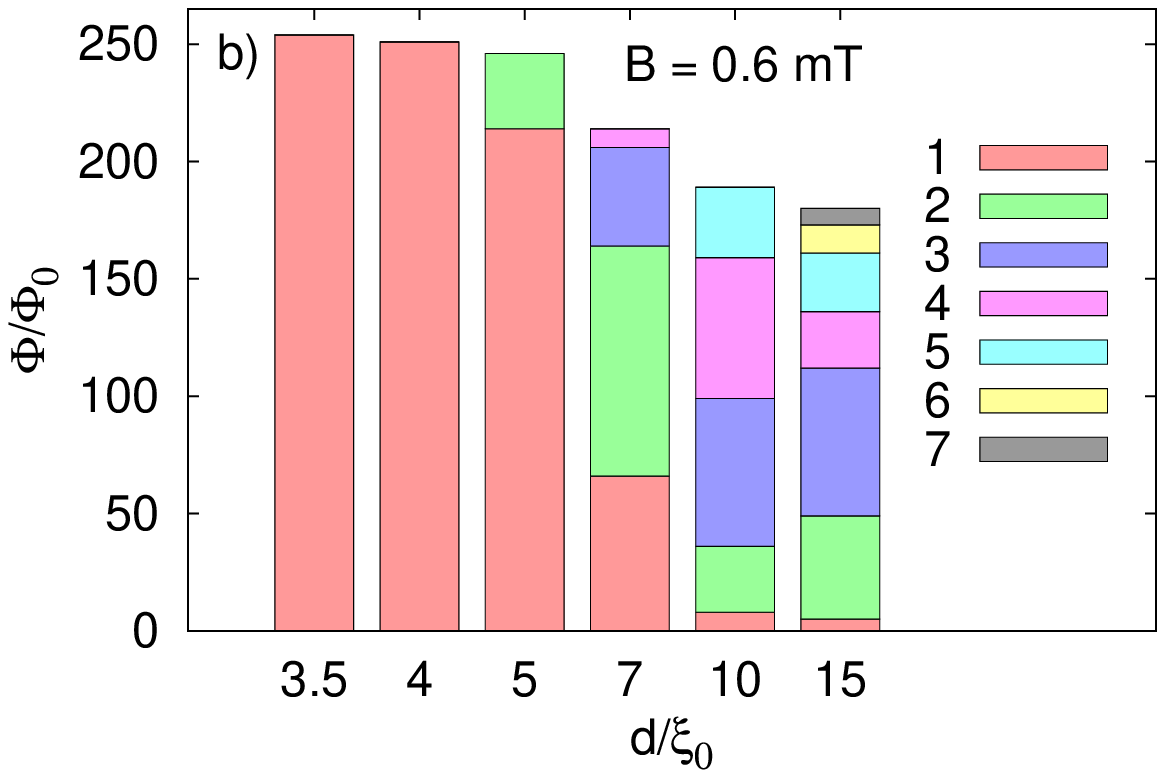}
\caption{The histograms show how the magnetic flux penetrating the sample is distributed between vortices of different vorticity that is indicated by the colours. The penetrating flux is shown as a function of magnetic field in the thickest film (a) and as a function of film thickness at the highest magnetic field (b). Both histograms are based on the simulations done with $\kappa = 0.5$.}
\label{histogrammi}
\end{center}
\end{figure}

In our work the giant vortices do not form a honeycomb lattice of vortices of the same vorticity as was predicted by Lasher~\cite{Lasher1}. Instead, giant vortices of different vorticities are mixed into the same lattice quite thoroughly, the smaller vortices filling in the gaps between the larger ones. The distributions of the giant vortices are shown as histograms in \fref{histogrammi} for the $\kappa = 0.5$ case. The thickest film with $d=15\xi_0$ is in the Meissner state at the lowest magnetic field. Just above the first critical field ($B=0.3$ mT) there are giant vortices with $n=2$ and a few with $n=1$. The one vortex with $n=4$ is sitting in a pinning site. Increasing the magnetic field brings in the higher vorticities up to $n = 7$. \Fref{histogrammi}b shows how the singly quantized vortices change into giant vortices when increasing the film thickness at $B=0.6$ mT. In the two thinnest films there are only singly quantized vortices present. The total flux decreases with increasing film thickness as expected from the change in the demagnetization factor with film thickness. The giant vortices seen in the simulations are a natural extension to the flux tube structure seen in the experiments on lead films in the intermediate state~\cite{Prozorov2}. The magnetic flux in a normal state spot in the intermediate state is quantized and a giant vortex is simply a very small flux tube in the intermediate state. Interestingly, the intermediate state continuously shrinks into a singly quantized vortex structure that is similar to the mixed state when the film thickness is decreased. The giant vortex structure serves as a transitional phase in between them.

\section{Conclusions}

We have simulated low-$\kappa$ superconducting thin films of different thicknesses. The results show that giant vortex structures are a stable solution to the Ginzburg-Landau equations even in low magnetic fields provided that the superconductor is of type I and of intermediate thickness. No complex geometries or pinning sites were needed for these giant vortices to form. On the contrary, the pinning sites broke the cylindrical symmetry of the giant vortices which turned them into multivortices. In the case of $\kappa=1$, no giant vortices were seen. Based on the distribution of the phase contours one can speculate that the reason behind the giant vortex state could be a short range attraction and long range repulsion between the flux lines. Smaller vortices were seen to coexist with giant vortices, the smaller ones filling the gaps. In addition, the size of the vortices was seen to increase with decreasing film thickness in a manner bearing some similarity with Pearl vortices. The results predict that singly quantized vortices could be stabilized in type I superconductors by the thin film geometry alone which was unexpected because the model does not take electron mean free path effects into account.

\ack
The authors thank CSC~--~Scientific Computing Ltd. for generous supercomputer resources. The Jenny and Antti Wihuri Foundation is acknowledged for financial support.


\section*{References}

\begin{thebibliography}{}

\bibitem{Hasegawa5}
{ Hasegawa S, Matsuda T, Endo J, Osakabe N, Igarashi M, Kobayashi T, Naito M,
  Tonomura A   and Aoki R }  1991 {\em Phys. Rev. B } {\bf 43} 7631

\bibitem{Cren2}
{ Cren T, Serrier-Garcia L, Debontridder F   and Roditchev D }  2011 {\em Phys.
  Rev. Lett.} {\bf 107} 097202

\bibitem{Engbarth1}
{ Engbarth M A, Bending S J   and Milo{\v s}evi{\'c} M V }  2011 {\em Phys.
  Rev. B } {\bf 83} 224504

\bibitem{Silhanek3}
{ Silhanek A V, Gutierrez J, Kramer R B G, Ataklti G W, {Van de Vondel} J,
  Moshchalkov V V   and Sanchez A }  2011 {\em Phys. Rev. B } {\bf 83} 024509

\bibitem{Kramer4}
{ Kramer R B G, Silhanek A V, {Van de Vondel} J, Raes B   and Moshchalkov V V }
   2010 {\em Physica C } {\bf 470} 758

\bibitem{Poole2}
{ {Poole Jr.} C P, Farach H A, Creswick R J   and Prozorov R }  2007 {\em
  Superconductivity, Second Edition}.
\newblock Academic Press

\bibitem{Tinkham2}
{ Tinkham M}  1963 {\em Phys. Rev.} {\bf 129} 2413

\bibitem{Maki5}
{ Maki K}  1965 {\em Ann. Phys. (N.Y.) } {\bf 34} 363

\bibitem{Lasher1}
{ Lasher G}  1967 {\em Phys. Rev.} {\bf 154} 345

\bibitem{Schweigert1}
{ Schweigert V A, Peeters F M   and Deo P S }  1998 {\em Phys. Rev. Lett.} {\bf
  81} 2783

\bibitem{Berdiyorov1}
{ Berdiyorov G R, Milo{\v s}evi{\'c} M V   and Peeters F M }  2006 {\em Physica
  C } {\bf 437-438} 25

\bibitem{Shi9}
{ Shi L M, Zhang L F, Meng H, Zhao H W, Zha G Q   and Zhou S P }  2009 {\em
  Phys. Rev. B } {\bf 79} 184518

\bibitem{Sweeney1}
{ Sweeney M C  and Gelfand M P }  2010 {\em Phys. Rev. B } {\bf 82} 214508

\bibitem{Palonen2}
{ Palonen H, J{\"a}ykk{\"a} J   and Paturi P }  2012 {\em Phys. Rev. B } {\bf
  85} 024510

\bibitem{Duan2}
{ Duan F  and Guojun J }  2005 {\em Introduction to Condensed Matter Physics,
  Volume 1}.
\newblock World Scientific Publishing, Singapore

\bibitem{Pearl1}
{ Pearl J}  1964 {\em Appl. Phys. Lett.} {\bf 5} 65

\bibitem{Ozer1}
{ {\"O}zer M M, Thompson J R   and Weitering H H }  2006 {\em Phys. Rev. B }
  {\bf 74} 235427

\bibitem{Cody1}
{ Cody G D  and Miller R E }  1966 {\em Phys. Rev. Lett.} {\bf 16} 697

\bibitem{Lischke1}
{ Lischke B  and Rodewald W }  1974 {\em Phys. Status Solidi B } {\bf 63} 97

\bibitem{5}
{ Brandt E H}  2005 {\em Phys. Rev. B } {\bf 71} 014521

\bibitem{Carneiro1}
{ Carneiro G  and Brandt E H }  2000 {\em Phys. Rev. B } {\bf 61} 6370

\bibitem{Clem5}
{ Clem J R}  1975 {\em J. Low Temp. Phys.} {\bf 18} 427

\bibitem{Lasher2}
{ Lasher G}  1965 {\em Phys. Rev.} {\bf 140} A523

\bibitem{Prozorov2}
{ Prozorov R, Giannetta R W, Polyanskii A A   and Perkins G K }  2005 {\em
  Phys. Rev. B } {\bf 72} 212508

\end{thebibliography}
\end{document}